\documentclass[12pt]{article}

\usepackage[dvips]{graphicx}

\def\sspace{\baselineskip = 0.24in}

\begin{document}
\sspace

\begin{titlepage}
\begin{flushright}
OSU-HEP-02-02\\
BA-02-04\\
February 2002\\
\end{flushright}
\vskip 2cm
\begin{center}
{\Large\bf Family Unification in Five and Six Dimensions}
\vskip 1cm
{\normalsize\bf
K.S.\ Babu$\,{}^1$, S.M. Barr$^2$, and Bumseok Kyae$^2$ }
\vskip 0.5cm
{\it ${}^1\,$Department of Physics, Oklahoma State University,\\
Stillwater, OK~~74078 \\ [0.15truecm]
${}^2\,$Bartol Research Institute, University of Delaware, \\Newark,
DE~~19716\\[0.1truecm]
}

\end{center}
\vskip .5cm

\begin{abstract}

In family unification models, all three families of quarks and leptons
are grouped together into an irreducible representation of a simple
gauge group, thus unifying the Standard Model gauge symmetries and
a gauged family symmetry. Large orthogonal
groups, and the exceptional groups $E_7$ and $E_8$ have been much studied 
for family unification. The main theoretical difficulty of family
unification is the existence of mirror families at the weak scale. 
It is shown here that family unification without mirror families
can be realized in simple five-dimensional and six-dimensional orbifold 
models similar to those recently proposed for $SU(5)$ and
$SO(10)$ grand unification. It is noted that a family unification group
that survived to near the weak scale and whose coupling extrapolated to 
high scales unified with those of the Standard model would be evidence 
accessible in principle at low energy of the existence of small (Planckian 
or GUT-scale) extra dimensions.

\end{abstract}
\end{titlepage}

\newpage

\section{Introduction}

\noindent
{\bf Family unification.}
\vspace*{0.1in}

It has long been known that theories with extra dimensions afford ways
of breaking gauge symmetries unavailable in $4d$ theories \cite{extrad}. 
Recently, such mechanisms
have been applied to the problem of constructing realistic grand unified
models \cite{kawamura, so10orb}. In particular, it has been shown that orbifold 
compactification of one or two extra dimensions allows symmetries to be 
broken in such a way as to resolve in an apparently simple fashion two of the
thorniest problems of $4d$ grand unification, namely the doublet-triplet
splitting problem and the problem of dimension-5
proton decay operators. In this
letter we show that the same ideas can be applied to a problem almost equally
old, that of ``family unification" \cite{ramond, wz, grs, o14, so16, so18, 
e7, e8}.

The idea of family unification is an extension of the idea of grand 
unification. Grand unification has two aspects, gauge unification and 
quark-lepton unification. Gauge unification is the unification 
of all the Standard Model gauge groups within a simple group, 
such as $SU(5)$, $SO(10)$, or $E_6$. Quark-lepton unification is the
idea that quarks and leptons are put together into irreducible 
multiplets of that simple group. There is complete quark-lepton unification
if all the fermions of one family are contained in a single irreducible 
representation. This is possible in $SO(10)$, $E_6$, and larger groups. 
The idea of family unification has the same two aspects. It involves,
first, the existence of a family gauge group that is unified with the 
Standard Model gauge group into a simple group \cite{ramond}, and, second,  
the existence of a single irreducible representation of that group
that contains the quarks and leptons of {\it all three} known 
families. For example, if there is a family group $SU(3)$, it can 
be unified with the Standard Model group within $E_8$, since 
$E_8 \supset E_6 \times SU(3) \supset SU(3) \times SU(2) \times U(1) 
\times SU(3)$. At the same time, in $E_8$ all three families can be 
unified into one irreducible representation, the ${\bf 248}$. The groups 
that have been most studied from the point of view of family unification 
are $O(14)$ \cite{o14}, $SO(16)$ \cite{so16}, $SO(18)$ \cite{so18}, 
$E_7$ \cite{e7}, and $E_8$ \cite{e8}. 

\vspace{0.2cm}

\noindent
{\bf The problem of mirror families.}
\vspace*{0.1in}

The central problem of family unification is that the ordinary $V-A$ 
families must be accompanied by an equal number of $V+A$ or ``mirror" 
families. For example, the ${\bf 248}$ of $E_8$ decomposes under
$E_6 \times SU(3)$ into
\begin{equation}
{\bf 248} \rightarrow ({\bf 78}, {\bf 1}) + ({\bf 1}, {\bf 8}) +
({\bf 27}, {\bf 3}) + (\overline{{\bf 27}}, \overline{{\bf 3}}).
\end{equation}

\noindent
Together with three ordinary families, $({\bf 27}, {\bf 3})$, one finds
three mirror families, $(\overline{{\bf 27}}, \overline{{\bf 3}})$.
Similarly, for the orthogonal groups:
a spinor of $SO(10 + 2n)$ decomposes under the $SO(10) \times SO(2n)$
subgroup as
\begin{equation}
2^{(4+n)} \rightarrow ({\bf 16}, 2^{(n-1)}) + 
(\overline{{\bf 16}}, 2^{(n-1) \prime}). 
\end{equation}

\noindent
(For $n$ odd the $2^{(n-1)}$ and $2^{(n-1) \prime}$ are conjugate to each
other, whereas for n even they are each self-conjugate. For a review
of spinor representations of orthogonal groups see \cite{wz}.) One sees that
for every ${\bf 16}$ there is a $\overline{{\bf 16}}$.

The fact that family unification generally gives equal numbers of families
and mirror families presents the problem of explaining why no
mirror families have been observed. The obvious solution would be to give 
large mass in some way to the mirror families. However, this turns out
to be not so easy. While mirror families could have large  
$SU(2)_L \times U(1)_Y$-invariant mass terms that couple them to ordinary 
families, such terms give mass to equal numbers of families and
mirror families and so do not resolve the problem. The only way to
give the mirror families large masses without giving large mass to the
ordinary families is with $SU(2)_L \times U(1)_Y$-breaking mass terms.
Such masses would have to be at or below the weak-interaction scale,
meaning that the mirror families should be observable through radiative
effects in precision tests of the electroweak theory. 
Recent analyses [12,13] conclude that some additional families
or mirror families are allowed, with Ref. [12] finding a chi--squared
minimum for the number of extra families to be somewhere between 
one and two.  Adler argues that even three mirror families may not
be completely excluded by present data [14].  However, even if mirror
families with masses in the hundred GeV range are consistent
with data, explaining in a natural way why their masses are much larger
than the observed quarks and leptons will be a theoretical challenge.  No
simple and plausible mechanism which attempts this is known.

Given that it is quite difficult to hide mirror families at the weak scale
the question arises whether in a model with family unification
the mirror families might be banished altogether from the low-energy spectrum.
One interesting possibility
is that the mirror families are confined at a high scale
\cite{grs, wz, gt}. Suppose, for example, that the family unification group is
$SO(18)$ and breaks down to $SO(10) \times SO(8)$ at a scale $M_G$.
The quarks and leptons are then in the representations
$({\bf 16}, {\bf 8}) + (\overline{{\bf 16}}, {\bf 8'})$. (See Eq. (2).)
If at a scale $M_{fam} \gg M_W$ the family group $SO(8)$ breaks to
some confining group $H$ under which the ${\bf 8'}$ decomposes into
only non-singlets, then all of the mirror families are confined. If, on
the other hand, ${\bf 8}$ contains some singlets in its decomposition
under $H$, then some ordinary families will be unconfined and appear
in the low-energy spectrum. For instance,
there is an $SO(5)$ subgroup of $SO(8)$ under which ${\bf 8'} \rightarrow
{\bf 4} + \overline{{\bf 4}}$ but ${\bf 8} \rightarrow {\bf 5} +
{\bf 1} + {\bf 1} + {\bf 1}$, so that if $SO(5)$ confined three
families would remain light \cite{grs, wz}. Unfortunately, it is doubtful 
that there exist models of family unification in four-dimensions in which 
the gauge coupling of $H$ is asymptotically free and thus able to 
confine the mirror families at some dynamically generated high scale. 
(The beta function for the $SO(5)$ gauge coupling in the example
given above is $d {g_5}/d{\rm ln}\mu = +\{(62/3)/(16 \pi^2)\} g_5^3$
in the non--supersymmetric case, with a larger positive value if
there is supersymmetry.) 

It would seem, therefore, that in four-dimensional theories mirror
families cannot be banished from the low-energy theory.
However, it can happen in theories with extra dimensions 
\cite{chsw, dhvw, gtgz}. For example,
as was shown in Ref. \cite{chsw}, in Calabi-Yau compactification of 
heterotic string theory, the $E_8$ of the observable sector can be broken 
down to $E_6$ below the compactification scale with a chiral low-energy 
spectrum of quarks and leptons. In Ref. \cite{dhvw} it was shown that 
orbifold compactification from ten to four dimensions can leave
an $E_6 \times SU(3)$ subgroup of $E_8$ unbroken, with a chiral low energy 
spectrum which contains some families in multiplets of the $SU(3)$.
(In the particular example of that paper there were thirty-six
families (in $3 \times ({\bf 27}, {\bf 3}) + 27 \times ({\bf 27}, {\bf 1})$)
and no mirror families.  
For early string theory models with three families, 
see Ref. \cite{3families}.) 
Such string theory models have family unification in the sense that all the 
families of quarks and leptons originate from one $E_8$ multiplet of the 
ten-dimensional theory. 

In this paper we show that family unification
can be achieved in simple $5d$ and $6d$ orbifold models quite
similar in spirit to those of \cite{kawamura, so10orb}. These are
``bottom up" models, like those of \cite{kawamura, so10orb}, and no
attempt is made to derive them from superstring theory.  In the next
section, we present 
simple illustrative models based on both orthogonal groups and $E_8$.
Before we turn to higher dimensions, however, it may be helpful to briefly 
review some facts about the family unification groups that have been studied 
in the literature.

\vspace{0.2cm}

\noindent
{\bf Family unification groups.}
\vspace*{0.1in}

First let us look at the smallest group that has been used for family
unification, $O(14)$. We see from Eq. (1) that
a spinor of $SO(14)$ decomposes under $SO(10) \times SO(4)$ as 
${\bf 64} \rightarrow ({\bf 16}, {\bf 2}) + (\overline{{\bf 16}}, 
{\bf 2}')$. The conjugate spinor decomposes as
$\overline{{\bf 64}} \rightarrow ({\bf 16}, {\bf 2}') + (\overline{{\bf 16}}, 
{\bf 2})$. Thus, a single spinor of $SO(14)$ can only accomodate two families
and their mirrors. However, if one takes the group to be $O(14)$ (that
is $SO(14)$ together with the parity that transforms the ${\bf 64}$
and $\overline{{\bf 64}}$ into each other) then the irreducible spinor
is ${\bf 128}$ ($= {\bf 64} + \overline{{\bf 64}}$), which can accomodate
four families. Models of family unification based on $O(14)$ have been 
constructed \cite{o14}.

The group $SO(16)$ is large enough for family unification, since
${\bf 128} \rightarrow ({\bf 16}, {\bf 4}) + 
(\overline{{\bf 16}}, \overline{{\bf 4}})$ (and
${\bf 128}' \rightarrow ({\bf 16}, \overline{{\bf 4}}) + 
(\overline{{\bf 16}}, {\bf 4})$). However, 
its spinors are real, i.e. self-conjugate, so that in the absence of any
other symmetry there would be nothing to prevent a mass term of the
form ${\bf 128} \cdot {\bf 128}$, which would naturally make all the quarks and 
leptons superheavy. (The same objection applies to $O(14)$.) For this reason
most authors have concentrated on the group $SO(18)$ \cite{so18}, which is the
smallest orthogonal group whose spinors are complex and large enough to
contain at least three families. However, it is possible to avoid the 
self-mass problem in $SO(16)$ by assuming that the ${\bf 128}$ of quarks 
and leptons is charged under some other group, say a $U(1)$. Interesting 
$SO(16)$ models using this idea have been constructed \cite{so16}.

The group $E_7$ has an adjoint that under the $SU(5) \times SU(3)$ subgroup
decomposes as ${\bf 133} \rightarrow ({\bf 24}, {\bf 1}) +
({\bf 1}, {\bf 8}) + ({\bf 5}, {\bf 1}) + (\overline{{\bf 5}}, {\bf 1})
+ ({\bf 1}, {\bf 1}) + ({\bf 10}, \overline{{\bf 3}}) + 
(\overline{{\bf 5}}, {\bf 3}) + (\overline{{\bf 10}}, {\bf 3}) +
({\bf 5}, \overline{{\bf 3}})$. As with the other groups we have discussed,
this has an equal number of families and mirror families. But it also
has the interesting peculiarity that some of the fermions in a family
transform as ${\bf 3}$ of the $SU(3)$ family group, while others
transform as $\overline{{\bf 3}}$. It should be noted also, that the
self-mass problem exists for both $E_7$ and $E_8$ (as it does for
$O(14)$ and $SO(16)$) since both the ${\bf 133}$ of $E_7$ and the
${\bf 248}$ of $E_8$ are real representations.

\section{Family Unification without the Mirror Families}

We start with a simple $SO(16)$ model of family unification in five 
dimensions in which the low-energy four-dimensional theory has
three light families in a triplet of $SU(3)$ and no mirror families.

Following in the footsteps of \cite{kawamura}, we consider a five-dimensional
theory with the fifth dimension ($y$) compactified on an $S^1/(Z_2 \times
Z_2')$ orbifold. The circumference of $S^1$ is $2 \pi R$. $Z_2$ reflects 
$y \rightarrow -y$, and $Z_2'$ reflects $y' \rightarrow - y'$, where 
$y' = y + \pi R/2$. (See Fig. 1.) The orbifold $S^1/(Z_2 \times Z_2')$
can be taken to be the interval $-\pi R/2 \leq y \leq 0$. The point
$y=0$, which we will call $O$, is a fixed point of $Z_2$, while the
point $y = -\pi R/2$ (or $y'=0$), which we will call $O'$, is a fixed point
of $Z_2'$. There are branes at the points $O$ and $O'$. 
It is assumed that
in the five-dimensional bulk there is an $N=1$ supersymmetric theory
with gauge group $SO(16)$ and fields consisting of a vector multiplet in
the adjoint (${\bf 120}$) and a hypermultiplet in the spinor (${\bf 128}$).

\begin{figure} 
\begin{center}
\includegraphics[width=60mm]{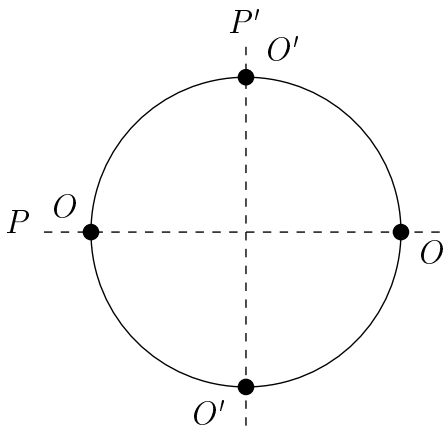}
\end{center}
\caption{
The fifth dimension compactified on an $S^1/(Z_2\times Z_2')$ orbifold.}
\end{figure}

The $N=1$ supersymmetry in five dimensions would give N=2 supersymmetry in
four dimensions, but the orbifold compactification will break this down
to $N=1$. The vector multiplet of the bulk theory splits into a vector
multiplet $V^a$ and a chiral multiplet $\Sigma^a$ of the unbroken $4d$ $N=1$
supersymmetry. Under the $Z_2$ symmetry $V^a$ is taken to
have parity $+$ and $\Sigma^a$ to have parity $-$. Similarly,
the hypermultiplet splits into two chiral multiplets $\Phi$ and 
$\Phi^c$ having opposite $Z_2$ parity. The $Z_2'$ parities are assigned
in the following way. The $SO(16)$ adjoint when decomposed under 
$SO(10) \times SO(6)$ gives ${\bf 120} \rightarrow ({\bf 45}, {\bf 1})
+ ({\bf 1}, {\bf 15}) + ({\bf 10}, {\bf 6})$. We assign $Z_2'$ parity
$+$ to the components of the $5d$ vector multiplet in the 
$({\bf 45}, {\bf 1}) + ({\bf 1}, {\bf 15})$, and $Z_2'$ parity $-$ to those 
in the
$({\bf 10}, {\bf 6})$. This assignment is consistent with the group algebra
of $SO(16)$. The spinor of $SO(16)$ decomposes as 
${\bf 128} \rightarrow ({\bf 16}, {\bf 4}) + (\overline{{\bf 16}},
\overline{{\bf 4}})$. Under $Z_2'$ we take the components of the $5d$ 
hypermuliplet that are in $({\bf 16}, {\bf 4})$ to have parity $+$ while 
those in $(\overline{{\bf 16}}, \overline{{\bf 4}})$ have parity $-$.
The complete $Z_2 \times Z_2'$ assignments are shown in Table I.
\vskip 0.6cm
\begin{center}
\begin{tabular}{|c||c|c|c|c|c|c|c|c|c|c|} \hline
 &$V_{({\bf 45},{\bf 1})}$ &$V_{({\bf 1},{\bf 15})}$ &$V_{({\bf 10},{\bf 6})}$&
$\Sigma_{({\bf 45},{\bf 1})}$&$\Sigma_{({\bf 1},{\bf 15})}$
&$\Sigma_{({\bf 10},{\bf 6})}$ &
$\Phi_{({\bf 16},{\bf 4})}$&$\Phi_{(\overline{{\bf 16}},\overline{{\bf 4}})}$&
$\Phi_{({\bf 16},{\bf 4})}^c$&
$\Phi_{(\overline{{\bf 16}},\overline{{\bf 4}})}^c$ \\ \hline \hline
$Z_2$ &+$$ &$+$ &$+$ &$-$ &$-$ &$-$ &$+$ &$+$ &$-$ &$-$ \\
$Z_2'$ &$+$ &$+$ &$-$ &$+$ &$+$ &$-$ &$+$ &$-$ &$-$ &$+$ \\ \hline
\end{tabular}
\vskip 0.4cm
{\bf Table I.~}($Z_2,Z_2'$) parity assignments for the
vector and hypermultiplets in $SO(16)$. 
\end{center}

As explained in many papers \cite{kawamura}, only those fields that
have $Z_2 \times Z_2'$ parity $(+,+)$ can have among their Kaluza-Klein
modes one that is constant in the fifth dimension, i.e. a zero mode
corresponding to a light field in the four-dimensional effective theory.
Fields that have $(+,-)$, $(-,+)$, or $(-,-)$ parities must vanish
at the fixed points $O$, $O'$, or both, and therefore have only higher
Kaluza-Klein modes, which are superheavy from the $4d$ point of view.
In this model the four-dimensional effective theory will have 
$N=1$ supersymmetry, gauge group $SO(10) \times SO(6)$, and light
fields coming from the bulk matter consisting of vector multiplets
in $({\bf 45}, {\bf 1})$ and $({\bf 1}, {\bf 15})$ and a chiral multiplet
in $({\bf 16}, {\bf 4})$. This cannot be the complete effective
low-energy theory since it has an $SO(6)^3$ gauge anomaly. One must
take into account also fields that ``live" on the branes at $O$ and $O'$.

On the brane at $O$ there is the full $SO(16)$ gauge symmetry, but on
the brane at $O'$ there is only an $SO(10) \times SO(6)$. It is not that
$SO(16)$ is spontaneously broken on the $O'$ brane, but rather that there
never was any gauge symmetry on $O'$ except $SO(10) \times SO(6)$. (At $O'$
quantities that are odd under $Z_2'$ have to vanish, including the gauge
parameters $\xi^a(x^{\mu}, y)$ for those generators $\lambda^a$ in
the coset $SO(16)/(SO(10) \times SO(6))$.) Consequently, in the $4d$ theory
on the brane at $O'$ only complete multiplets of $SO(10) \times SO(6)$ but 
not $SO(16)$ need appear. Let us suppose then
that on the $O'$ brane one has quark and lepton chiral multiplets
in $(\overline{{\bf 16}}, {\bf 1}) + 2 \times ({\bf 1}, \overline{{\bf 10}})$
and Higgs chiral multiplets in $({\bf 1}, {\bf 4}) + 
({\bf 1}, \overline{{\bf 4}})$. (We will denote the Higgs multiplets 
henceforth with subscript `$H$'.) The low-energy theory then has
altogether the following chiral multiplets:
$({\bf 16}, {\bf 4}) + (\overline{{\bf 16}}, {\bf 1}) + 
2 \times ({\bf 1}, \overline{{\bf 10}}) + ({\bf 1}, {\bf 4}) + 
({\bf 1}, \overline{{\bf 4}})$. This set has no gauge anomalies. 
(The $SO(6)^3$ anomaly of a $\overline{{\bf 10}}$ is $-8$ times that of
a ${\bf 4}$, so the total $SO(6)^3$ anomaly of the above set is
$16(1) + 2(-8) = 0$.) 

We assume that the $({\bf 1}, {\bf 4})_H + ({\bf 1}, \overline{{\bf 4}})_H$
obtain vacuum expectation values at some scale $M_{family} \gg M_W$
that breaks $SO(6)$ ($= SU(4)$) down to $SU(3)$. Then a term of the
form $({\bf 16}, {\bf 4}) (\overline{{\bf 16}}, {\bf 1}) 
({\bf 1}, \overline{{\bf 4}})_H$ on the $O'$ brane
will give mass to the $(\overline{{\bf 16}}, {\bf 1})$ 
$V+A$ family and one of the four $V-A$ families to
leave three light families in a $({\bf 16}, {\bf 3})$ of
$SO(10) \times SU(3)$. Note that the scale $M_{family}$ can be anything
from the weak scale up to the compactification scale. 

There still remain the questions of how $SO(10)$ breaks down to the 
Standard Model group in a satisfactory way, how the electroweak
symmetry is broken, how realistic masses arise for the quarks and leptons,
and how the $SU(3)$ family symmetry breaks. These breakings can
all be accomplished through the ordinary Higgs mechanism by Higgs
multiplets living on the $O'$ brane in suitable representations
of $SO(10) \times SO(6)$. We know, for example, that breaking of
$SO(10)$ down to the Standard Model can be achieved in $4d$ SUSY GUTs
in such a way as to have natural doublet-triplet splitting \cite{dw}
and also suppression of dimension-5 proton decay operators (i.e. those coming
from colored Higgsino exchange) \cite{bb}. It would be interesting to
attempt to use orbifold compactification to break $SO(10)$ down to the
Standard Model as well, as in \cite{so10orb}. However, we do not pursue 
that more ambitious goal in this paper.

What we have shown in this simple example is that models can be constructed
in which the three light families come from a single irreducible
representation of a simple group --- here a hypermultiplet in the bulk
$SO(16)$ theory, and in which that simple group comprises both the Standard
Model gauge symmetries and a family gauge symmetry that survives in the
four-dimensional theory, perhaps even down to low energy. 
It is clear that one may construct $SO(18)$ models in a similar way.
However, the
analogous orbifold breaking to $SO(10) \times SO(8)$ would
give {\it eight} families coming from the hypermultiplet in the bulk. 
One would therefore have to introduce {\it five} mirror families on 
the $O'$ brane, and a correspondingly
large number of Higgs on the brane to mate the mirror families with the
ordinary ones. It would be interesting to know if a more economical
and compelling way to obtain three light families from $SO(18)$
using orbifold compactifications could be found.

We see, then, that the orthogonal groups can accomodate three light families,
but they do not naturally prefer three light families. If we wish to
explain the fact that there are three light families, the group $E_8$
seems more promising, given that $E_6 \times SU(3)$ is a maximal
subgroup and that the fundamental representation of $E_8$ contains
a $({\bf 27}, {\bf 3})$ under that subgroup, as shown in Eq. (1).

It is easy to see that orbifolds with $Z_2$ symmetries as we have been
considering up to now are not adequate
to split the $(\overline{{\bf 27}}, \overline{{\bf 3}})$ mirror
families from the $({\bf 27}, {\bf 3})$ families in the ${\bf 248}$.
For a $Z_2$ parity to accomplish this, the 
$(\overline{{\bf 27}}, \overline{{\bf 3}})$ and $({\bf 27}, {\bf 3})$
would have to have opposite parity. Suppose, then, we imagine the
parity of $({\bf 27}, {\bf 3})$ to be $+$ and that of
$(\overline{{\bf 27}}, \overline{{\bf 3}})$ to be $-$. This would be
inconsistent with the group algebra of $E_8$. The 248 generators of $E_8$
fall into $({\bf 78}, {\bf 1}) + ({\bf 1}, {\bf 8}) +
({\bf 27}, {\bf 3}) + (\overline{{\bf 27}}, \overline{{\bf 3}})$.
The commutator of two $({\bf 27}, {\bf 3})$ generators gives a
$(\overline{{\bf 27}}, \overline{{\bf 3}})$ generator. Similarly,
the commutator of two $(\overline{{\bf 27}}, \overline{{\bf 3}})$ generators
gives a $({\bf 27}, {\bf 3})$ generator. Consequently the gauge fields
in $({\bf 27}, {\bf 3})$ and $(\overline{{\bf 27}}, \overline{{\bf 3}})$
must both have $+$ parity. That implies that the families and mirror families
have the same parity.

In order to eliminate the mirror families from the $4d$ theory in $E_8$
there must be an orbifold with at least a $Z_3$ symmetry. The point is
that under a $Z_3$ the generators of $E_8$ can transform as follows
consistently with the group algebra:

\begin{equation}
\begin{array}{lcl}
\lambda_{(78,1)} & \longrightarrow & \lambda_{(78,1)} \\ & & \\
\lambda_{(1,8)} & \longrightarrow & \lambda_{(1,8)} \\ & & \\
\lambda_{(27,3)} & \longrightarrow & \omega \; \lambda_{(27,3)} \\ & & \\
\lambda_{(\overline{27}, \overline{3})} & \longrightarrow &
\omega^2 \; \lambda_{(\overline{27}, \overline{3})}, \\ & & 
\end{array}
\end{equation}

\noindent
where $\omega = e^{2\pi i/3}$.
Such a $Z_3$ can be used to split the families from the mirror families 
in a ${\bf 248}$, as will be seen.

Consider a six-dimensional theory in which the extra two dimensions
($x^5$ and $x^6$) are compactified on an orbifold with a $Z_3$
symmetry. The construction of such an orbifold is shown in Fig. 2. 
One starts with a torus $T^2$ defined by identifying the points 
$z \equiv x^5 + i x^6$ with the points $z + 1$ and $z + \omega$.
This torus is shown as the large parallelogram in Fig. 2. Under a
rotation by $2 \pi /3$ (i.e. $z \rightarrow \omega z$)
one sees that the regions labelled 1 get mapped into other regions 
labelled 1, and that similarly 2 gets mapped into 2, 3 into 3, and 4 into 4.
The $T^2/Z_3$ orbifold obtained from $T^2$ by identifying $z$ with 
$\omega z$ 
thus has as its fundamental region the smaller parallelogram $aba'c$.
This orbifold has a $Z_3$ symmetry, under which $a$, $b$, and $c$
are inequivalent fixed points. (The point $a'$ is identified with $a$).
(The orbifold used in \cite{dhvw} is the six-dimensional generalization
of this simple two-dimensional orbifold and has 27 fixed points.)

\begin{figure}[t]
\label{bps}
\begin{center}
\includegraphics[width=60mm]{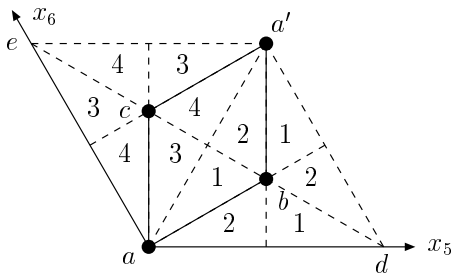}
\end{center}
\caption{ 
The two extradimensions $x_5$ and $x_6$ compactified on an $S^1/Z_3$
orbifold.  The regions with the same numbers
are identified each other.
$a$, $b$, and $c$ are fixed points of $Z_3$, and the parallelogram $aba'c$ 
is the fundamental region.  
}
\end{figure}

Instead of using one discrete symmetry to break supersymmetry down to
$N=1$ and another discrete symmetry to break the gauge group as before,
we will use the same $Z_3$ to break both.
In the six-dimensional bulk we take the theory to have $N=1$ supersymmetry
and gauge group $E_8$. (One need not worry about cancellation of
gauge anomalies since the $E_8$ box anomaly in six dimensions
vanishes up to ``reducible" pieces that can be cancelled by the
Green-Schwarz mechanism \cite{anomaly}.) We take the matter in the bulk 
to consist of a vector multiplet in the ${\bf 248}$. 
A $6d$ vector multiplet decomposes, under the $4d$ $N=1$ supersymmetry, into 
a vector multiplet $V^a$ and a chiral multiplet $\Sigma^a$.
The index $a$ labels the 248 generators of $E_8$, which may be
classified under the $E_6 \times SU(3)$ subgroup as shown in Eq. (1).
The fields in $V^a$ are assumed to transform under $Z_3$ in the same way
as the corresponding generators, which is shown in Eq. (3).
The fields in $\Sigma^a$ transform in the same way except with an additional
factor of $\omega^2$. The $Z_3$ transformation properties of the fields 
are given in Table II. 

\vskip 0.6cm
\begin{center}
\begin{tabular}{|c||c|c|c|c|c|c|c|c|} \hline
 &$V_{({\bf 78},{\bf 1})}$ &$V_{({\bf 1},{\bf 8})}$ &$V_{({\bf 27},{\bf 3})}$&
$V_{(\overline{{\bf 27}},\overline{{\bf 3})}}$ &
$\Sigma_{({\bf 78},{\bf 1})}$&$\Sigma_{({\bf 1},{\bf 8})}$
&$\Sigma_{({\bf 27},{\bf 3})}$ &
$\Sigma_{(\overline{{\bf 27}},\overline{{\bf 3}})}$ \\ \hline \hline
$Z_3$ &1$$ &$1$ &$\omega$ &$\omega^2$ &$\omega^2$ &$\omega^2$ &$1$ &$\omega$
\\ \hline
\end{tabular}
\vskip 0.4cm
{\bf Table II.~}$Z_3$ transformation properties of the vector and chiral
multiplets in $E_8$.
\end{center} 

Only those component fields that transform trivially under $Z_3$ will
have zero modes. Thus, the bulk matter contributes the following fields 
to the $4d$ low energy theory: vector multiplets in $({\bf 78}, {\bf 1})$
and $({\bf 1}, {\bf 8})$ containing the gauge fields and
gaugino fields of $E_6$ and $SU(3)$, and a chiral multiplet in 
$({\bf 27}, {\bf 3})$ containing three families of quarks and leptons.
This cannot be the whole story as the $4d$ $SU(3)^3$ gauge anomaly must be 
cancelled. This can be done by chiral multiplets living on one or more of the
branes at the fixed points $a$, $b$, and $c$. For example, the
$SU(3)^3$ anomaly would be cancelled by twenty-seven 
$({\bf 1}, \overline{{\bf 3}})$ or by a single $({\bf 1}, 
\overline{{\bf 10}})$ of $E_6 \times SU(3)$ on one of the branes. 
It should be noted that on the branes there is only an $E_6 \times SU(3)$
symmetry to start with, so that it is only necessary to have complete
multiplets of that group rather than of the bulk gauge group $E_8$.
Again, as in the previous example, we can introduce various (vectorlike)
Higgs representations on the branes to break the symmetry down to that of 
the Standard Model. Such model-building details are beyond the scope of 
this paper. But we see no obstacle to constructing a fully realistic model
in this way.

\section{Family Unification as an indicator of Extra Dimensions}

As we have seen, family unification involves the idea that all three
known families reside within one irreducible representation of some
simple group. This group contains not only the Standard Model gauge groups,
but generally a family gauge group. In many cases this family gauge group
contains an $SU(3)$ subgroup (or is an $SU(3)$) under which the three
families form a triplet. The family gauge group could be broken at
very high scales, either by an ordinary $4d$ Higgs field or by the
orbifold compactification. But it is also quite possible that a
local family symmetry, perhaps an entire $SU(3)$,
 survives down to ``low scale", by which we mean
here something in the hundred $TeV$ range.   

If family unification is realized in Nature and the family group survives
down to low energy the interesting possibility would exist, at least in 
principle even if quite difficult in practice,
to infer the existence of near-Planckian extra dimensions from 
experiments done at ``low energy". 

Let us call the set of representations of $G_{SM} = SU(3)_c \times SU(2)_L
\times U(1)_Y$ that make up one family of left-handed fermions ${\bf F}$,
and suppose for specificity that the low energy family symmetry is
$SU(3)$ (though most of what we will say applies as well to other
groups, such as $SO(3)$).
Typically, in models with family unification the known quarks and leptons 
are in $({\bf F}, {\bf 3})$ of $G_{SM} \times SU(3)_{fam}$. 
(It is also
possible that some of the left-handed fermions of a family are in 
${\bf 3}$ and some in $\overline{{\bf 3}}$ of $SU(3)_{fam}$, as we saw in
the case of family unification in the adjoint of $E_7$. What we say
below applies to such cases as well.) 

If $SU(3)_{fam}$ is broken at a scale $M_{fam}$ near enough to the weak 
scale to be accessible to experiment, then one might be able eventually
to do two things: first, to find out what representations of the family group
the quarks and leptons are in; and, second, to measure the $SU(3)_{fam}$ 
gauge coupling $g_{fam}$ and extrapolate it to high energies to see whether 
it unifies with the Standard Model couplings. 
Suppose that this is done and it is found that 
the quarks and leptons of the Standard Model are indeed in 
$({\bf F}, {\bf 3})$, with no light mirror families, and that the 
four gauge couplings do indeed unify. What would one be able
reasonably to infer from this?

First, the evidence for the unification of the four gauge groups
(into one simple group or a product of identical simple groups related by
a discrete symmetry) would then be compelling. The unification that we now see
of the three Standard Model gauge couplings might be an accident, since
it really involves only one non-trivial condition being satisfied. However,
the accidental unification of {\it four} couplings at one point could
hardly be dismissed as accidental.

One would therefore conclude that at a high scale some unified group
$G$ broke down to $G_{SM} \times SU(3)_{fam}$, and that some
anomaly free set of representations of $G$ broke into
$({\bf F}, {\bf 3})$ plus some additional multiplets that were vectorlike under
$G_{SM}$. (They would have to be vectorlike under $G_{SM}$ in order
not to have been seen at the weak scale.) The question that would then 
confront theorists is whether such a breaking could take place in the
context of a theory with only four spacetime dimensions. 
The answer seems to be no.

We have seen that groups such as $SO(16)$, $SO(18)$, and $E_8$
have representations that contain $({\bf F}, {\bf 3})$ of 
$G_{SM} \times SU(3)$, but these are always accompanied by mirror families, 
and in four dimensions there does not seem to be any way to push these
mirror families to very large scales. The unitary groups, in particular
$SU(N)$ with $N \geq 8$, also can break down to $G_{SM} \times SU(3)$ giving
Standard Model multiplets that are in triplets of $SU(3)$, but
one never gets simply three families plus pieces that are
vectorlike under $G_{SM}$. For example consider $SU(8)$ with quarks
and leptons in $\psi^{[\alpha \beta \gamma]} + \psi_{[\alpha \beta]}
+ \psi_{\alpha}$. Under the $SU(5) \times SU(3)$ subgroup these do give
$({\bf 10}, {\bf 3}) + (\overline{{\bf 5}}, \overline{{\bf 3}})$, but
these come together with 
$2 (\overline{{\bf 10}}, {\bf 1}) + ({\bf 5}, \overline{{\bf 3}})
+ (\overline{{\bf 5}}, {\bf 1}) + ({\bf 1}, {\bf 3}) +
({\bf 1}, \overline{{\bf 3}}) + ({\bf 1}, {\bf 1})$, which are chiral under
$G_{SM}$.

The closest that one seems to be able to come is with the group
$SU(5) \times SU(5)$, with the two gauge couplings forced to be equal
by a discrete symmetry that interchanges the two $SU(5)$'s.
Consider the following anomaly-free set of fermions:
$({\bf 10}, \overline{{\bf 5}}) + (\overline{{\bf 5}}, {\bf 10})
+ 2 (\overline{{\bf 10}}, {\bf 1}) + 2 ({\bf 1}, \overline{{\bf 10}})
+ 7 ({\bf 5}, {\bf 1}) + 7 ({\bf 1}, {\bf 5})$.
Under the $SU(5) \times SU(3)$ subgroup this gives:
$({\bf 10}, \overline{{\bf 3}}) + (\overline{{\bf 5}}, \overline{{\bf 3}})$ 
together with
$2 (\overline{{\bf 5}}, {\bf 3}) + 6({\bf 5}, {\bf 1})$ and other 
pieces that are vectorlike under $SU(5) \times SU(3)$. This indeed
gives $({\bf F}, \overline{{\bf 3}})$ plus pieces that are vectorlike
under $G_{SM}$. However, as one can see, those pieces include {\it six}
$\overline{{\bf 5}} + {\bf 5}$ of $SU(5)$ that cannot get mass
above $M_{fam}$ because they are chiral under $SU(3)_{fam}$. 
In a supersymmetric theory this would make the couplings
blow up below the unification scale.

On the other hand, we have seen that in theories with extra dimensions
one can end up with three families in triplets of an 
$SU(3)$ family gauge symmetry whose coupling unifies at high scales with
the Standard Model gauge couplings, without there being any light mirror 
families. Such a situation, if it is found, would be a telltale sign 
at low energy of a theory with extra dimensions at very high scales
not directly accessible to experiment.

\vspace*{0.1in}
\noindent {\bf Acknowledgments} 
\vspace*{0.1in}

The work of KSB is supported in part by DOE Grant $\sharp$ DE-FG03-98ER-41076, 
a grant from the Research Corporation and by 
DOE Grant $\sharp$ DE-FG02-01ER-45684. 
The work of SMB and BK is supported in part by 
DOE Grant $\sharp$ DE-FG02-91ER-40626.

\end{document}